\documentclass{SciPost}

\hypersetup{
    colorlinks,
    linkcolor={red!50!black},
    citecolor={blue!50!black},
    urlcolor={blue!80!black}
}

\usepackage[bitstream-charter]{mathdesign}
\DeclareSymbolFont{usualmathcal}{OMS}{cmsy}{m}{n}
\DeclareSymbolFontAlphabet{\mathcal}{usualmathcal}

\fancypagestyle{SPstyle}{
\fancyhf{}
\lhead{\colorbox{blue}{\bf \color{white} ~Proceedings }}
\rhead{{\bf \color{scipostdeepblue} ~Klaus WERNER }}

\fancyfoot[C]{\textbf{\thepage}}
}

\begin{document}

\pagestyle{SPstyle}

\vspace*{0.2cm}  

\begin{center}{\Large \textbf{\color{scipostdeepblue}{
EPOS4: New theoretical concepts for modeling proton-proton and ion-ion scattering at very high energies  \\
}}}\end{center}

\begin{center}\textbf{
Klaus Werner\textsuperscript{1}
}\end{center}

\begin{center}
{\bf 1} SUBATECH, Nantes University -- IN2P3/CNRS -- IMT Atlantique, Nantes, France
\end{center}

\definecolor{palegray}{gray}{0.95}
\begin{center}
\colorbox{palegray}{
  \begin{tabular}{rr}
  \begin{minipage}{0.36\textwidth}

  \end{minipage}
  &
  \begin{minipage}{0.55\textwidth}
    \begin{center} \hspace{5pt}
    {\it 22nd International Symposium on Very High \\Energy Cosmic Ray Interactions (ISVHECRI 2024)} \\
    {\it Puerto Vallarta, Mexico, 8-12 July 2024} \\
    \end{center}
  \end{minipage}
\end{tabular}
}
\end{center}

\section*{\color{scipostdeepblue}{Abstract}}
\textbf{\boldmath{%
I explain the new concepts underpinning EPOS4, a novel theoretical framework designed to model hadronic interactions at ultrarelativistic energies. This approach eventually reconciles the parallel multiple scattering scenario (needed in connection with collective effects) and factorization (being the conventional method for high-energy scattering).
}}

\vspace{\baselineskip}




\vspace{10pt}
\noindent\rule{\textwidth}{1pt}
\tableofcontents
\noindent\rule{\textwidth}{1pt}
\vspace{10pt}


\section{Introduction}
\label{sec:intro}

\begin{figure}
\noindent \begin{centering}
\includegraphics[bb=0bp 0bp 713bp 450bp,clip,scale=0.32]{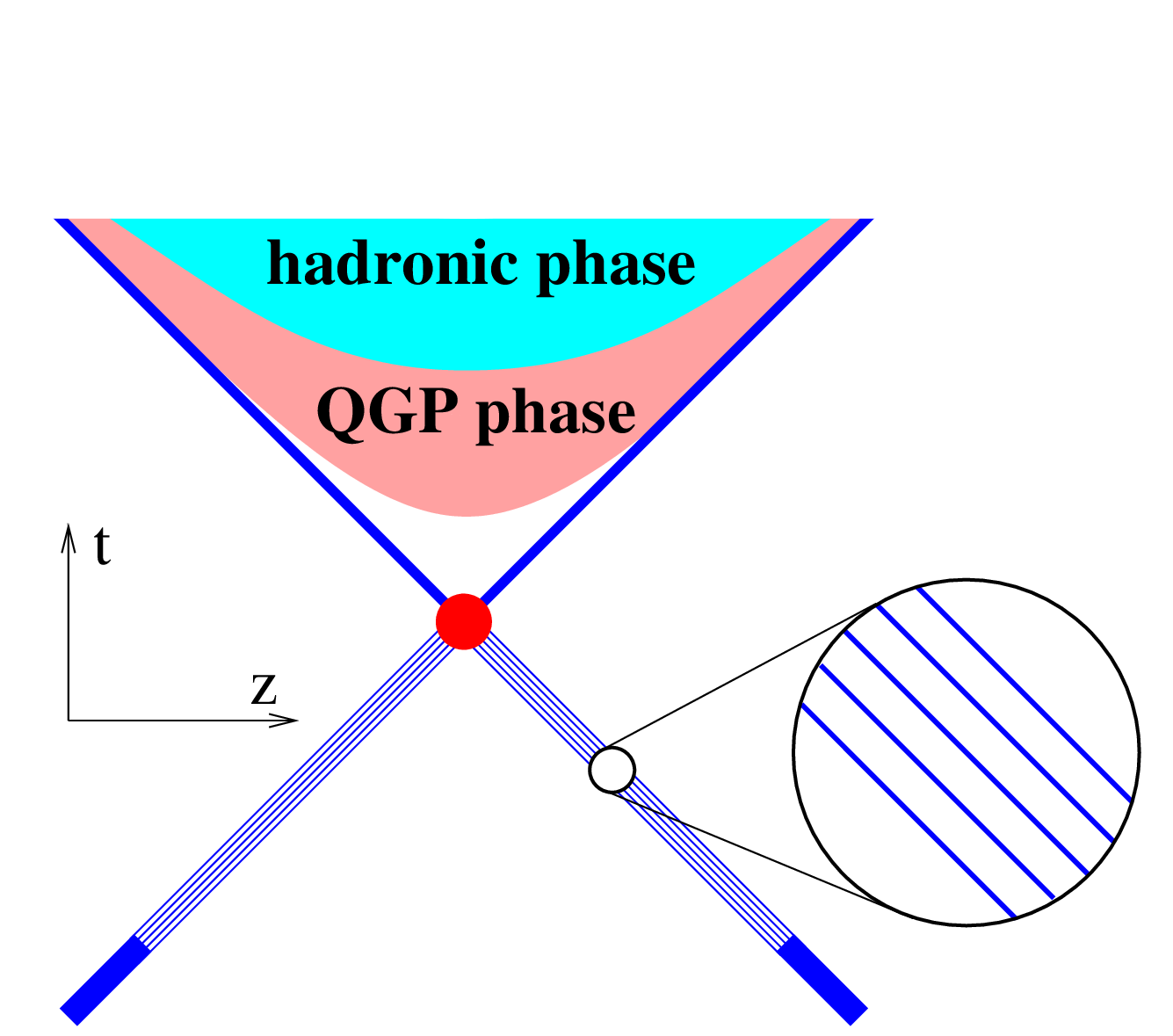}
\par\end{centering}
\caption{Space-time picture of hadronic scatterings.\label{fig:Space-time-picture}}

\end{figure}

In Fig. \ref{fig:Space-time-picture}, one can see the typical space-time representation of high-energy hadronic scatterings. The process begins with primary interactions occurring within a pointlike overlap zone (depicted as a red point) in proton-proton ($pp$) collisions, as well as in proton-nucleus (\emph{pA}) or nucleus-nucleus (\emph{AA}) collisions. Subsequently, the formation of the quark-gluon plasma (QGP) and the production of final state hadrons occur at a later stage.

In the diagram, it is evident that a comprehensive representation of space-time must consider the prior splitting of partons (parton evolution). This process (which prepares the actual scattering) takes a long time due to significant $\gamma$ factors. However, the interaction region (depicted in red) is indeed pointlike, necessitating multiple scatterings to occur simultaneously (the long "preparation" does not allow to have the scatterings one after the other). In the EPOS4 approach for primary interactions, one avoids sequential scatterings for both parton-parton and nucleon-nucleon interactions by rigorously conducting multiple scatterings in parallel. This is true for both, the theoretical formalism and the Monte Carlo realization, based on the principle that the Monte Carlo must be derived directly from theory, which is a non-trivial task.

The EPOS4 method has been previously introduced in a series of technical papers \cite{werner:2023-epos4-overview,werner:2023-epos4-heavy,werner:2023-epos4-smatrix,werner:2023-epos4-micro} that span 160 pages and aim to provide comprehensive details to avoid treating EPOS4 as a black box. These papers address numerous solved technical challenges, such as $N$-dimensional integrals and probability laws with $N>10^{6}$. However, beyond these technical aspects lie new and distinctive concepts that need to be clearly elucidated, along with explanations of their functionality. This is the main objective of this communication. These concepts establish a connection between pre-QCD multiple scattering approaches \cite{Gribov:1967vfb,Gribov:1968jf,GribovLipatov:1972,Abramovskii:1973fm} and the standard tool in high-energy scattering, the factorization approach \cite{Collins:1989,Ellis:1996}.

Some technical remarks: I use the symbols $\boldsymbol{\mathrm{T}}$, $T$, $G$,
and $\tilde{\sigma}$ as explained in Tab. \ref{tab:Important-symbols},
where I use transverse nucleon coordinates $b_{i}^{A}$ and $b_{j}^{B}$,
and the nuclear thickness function $T_{A}(b)=\int dz\,\rho_{A}\left(\sqrt{b^{2}+z{}^{2}}\right)$,
where $\rho_{A}$ is the (normalized) nuclear density for the nucleus
$A$. 

\noindent 
\begin{table}
\noindent %
\begin{tabular}{|l|l|}
\hline 
 & meaning\tabularnewline
\hline 
\hline 
$\boldsymbol{\mathrm{T}}(s,t)$ & %
\begin{minipage}[t]{0.75\columnwidth}%
elastic scattering T-matrix;\\
 $s$, $t$ Mandelstam variables%
\end{minipage}\tabularnewline
\hline 
$T(s,b)$ & %
\begin{minipage}[t]{0.75\columnwidth}%
Fourier transformation of $\boldsymbol{\mathrm{T}}(s,t)$ with respect
to the momentum transfer, divided by $2s$ (impact parameter representation)%
\end{minipage}\tabularnewline
\hline 
$G$ & %
\begin{minipage}[t]{0.75\columnwidth}%
2 Im$T$ \textendash{} representing inelastic scattering (cut diagram)%
\end{minipage}\tabularnewline
\hline 
$\tilde{\sigma}$ & %
\begin{minipage}[t]{0.75\columnwidth}%
Integrand in cross section formulas:\\
$pp$: ~~~~~ $\sigma^{pp}=\int d^{2}b\,\tilde{\sigma}^{pp}(s,b)$
\emph{}\\
\emph{A}+\emph{B}:~~~ $\sigma^{AB}=\int db\!_{A\!B}\ \tilde{\sigma}^{AB}(s,b,\{b_{i}^{A}\},\{b_{i}^{B}\})$\\
with\\
~~~~ $\int db\!_{A\!B}=\int d^{2}b\int\prod_{i=1}^{A}d^{2}b_{i}^{A}\:T_{A}(b_{i}^{A})$\\
\hspace*{2cm}$\int\prod_{j=1}^{B}d^{2}b_{j}^{B}\:T_{B}(b_{j}^{B})$%
\end{minipage}\tabularnewline
\hline 
\end{tabular}
\caption{Important symbols.\label{tab:Important-symbols}}
\end{table}

\begin{figure}
\begin{centering}
\includegraphics[scale=0.32]{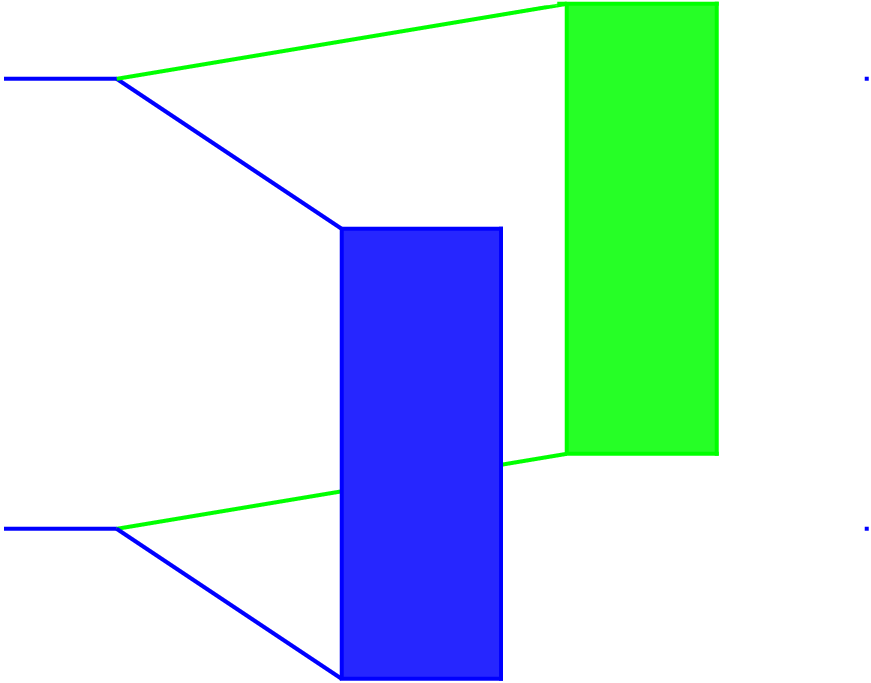}
\par\end{centering}
\caption{Double scattering in GR.\label{fig:Double-scattering}}
\end{figure}

Let us start by delving into the Gribov-Regge (GR) approach, as documented in \cite{Gribov:1967vfb,Gribov:1968jf,GribovLipatov:1972,Abramovskii:1973fm}. In this approach, multiple scattering in $pp$ occurs in a strictly parallel manner, as illustrated in Fig. \ref{fig:Double-scattering}. Each box represents an inelastic subscattering $G$, leading to chains of particles, with the specific mechanism being unknown at that time. There is no conflict even when a lengthy "preparation" is required, as discussed earlier. In the GR approach, cross sections are expressed in terms of weights $P$ that depend on the single scattering expression $G$ as
\begin{equation}
\tilde{\sigma}_{\mathrm{in}}^{pp}\!=\!\!\sum_{m=1}^{\infty}\underbrace{\frac{1}{m!}G^{m}\,e^{-G}}_{P(m)},\ \tilde{\sigma}_{\mathrm{in}}^{AB}\!=\!\!\sum_{\{m_{k}\}}\underbrace{\prod_{k=1}^{AB}\mathrm{\frac{1}{m_{k}!}}(G_{k})^{m_{k}}\ e^{-G_{k}}}_{P(\{m_{k}\})}
\end{equation}
where the terms $P(m)$ and $P(\{m_{k}\})$, representing probability
distributions, may serve as a basis for Monte Carlo applications. Let
me discuss, step by step, the improvements realized in the EPOS4 approach.

\section{Adding energy-momentum conservation}
\label{sec1}

\begin{figure}
\begin{centering}
\includegraphics[scale=0.32]{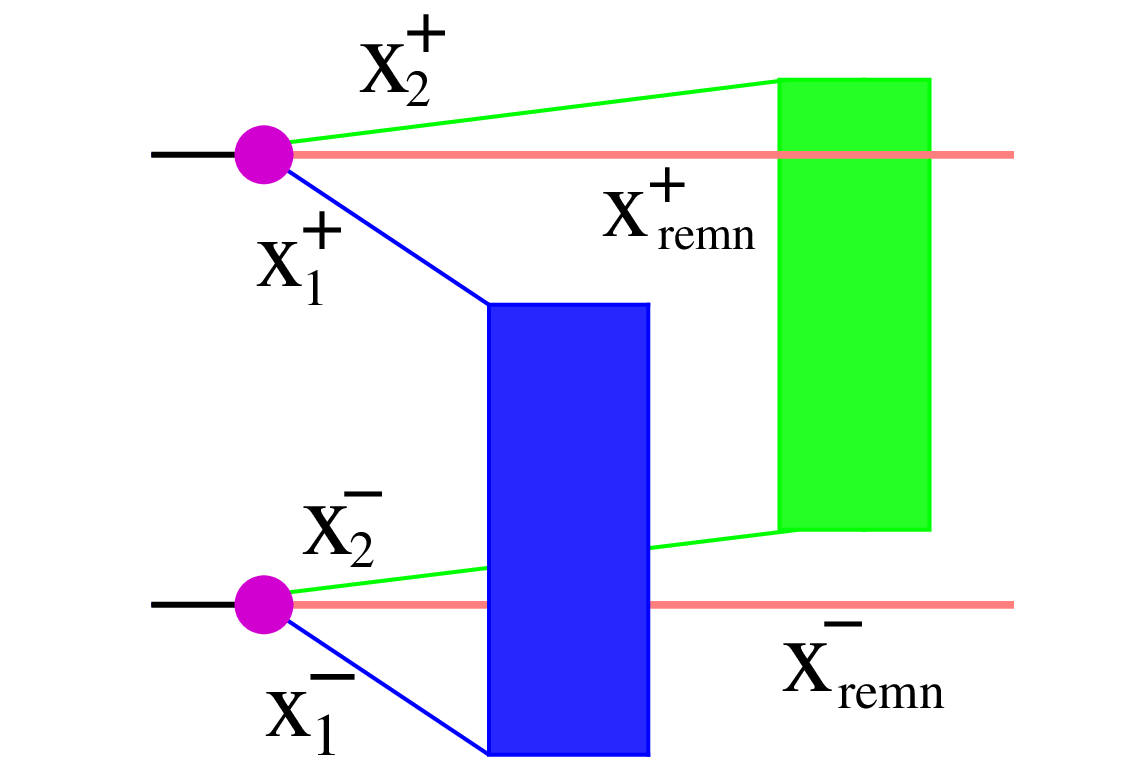}
\par\end{centering}
\caption{Energy-momentum sharing (GR\protect\textsuperscript{+}) in EPOS4.\label{fig:Energy-momentum-sharing}}

\end{figure}

In some cases, energy-momentum conservation is not particularly significant (such as for total cross sections), but for other cases, it is absolutely essential (like in particle production).
It is also necessary as a solid basis for Monte Carlo applications.
As discussed in detail in Ref. \cite{werner:2023-epos4-smatrix}:  To ensure energy-momentum sharing (GR\textsuperscript{+}) in EPOS4,
in \emph{pp} or for each \emph{NN} scattering in \emph{A}+\emph{B,
}one considers (compared to GR) new variables: the lightcone momentum
fractions $x_{m}^{+}$ and $x_{m}^{-}$ of subscatterings, with
\begin{equation}
 x_{\mathrm{remn}}^{\pm}=1-\!\!\sum x_{m}^{\pm}\ ,
\end{equation}
being the lightcone momentum fraction of the remnant, see Fig. \ref{fig:Energy-momentum-sharing}.

The expressions for cross sections, as shown in Ref.  \cite{werner:2023-epos4-smatrix}, still use weights \emph{P(K)} for configurations
\begin{equation}
K=\big\{\{m_{k}\},\{x_{k\mu}^{\pm}\}\big\},
\end{equation}
referring to $m_{k}$
subscatterings per pair $k$, with lightcone momentum fractions $x_{k\mu}^{\pm}$. 

This provides a solid basis for
Monte Carlo simulations: one determines $K$ according to $P(K)$,
instantaneously, there are no sequences, everything happens in parallel.
And one has MC = theory.

\section{Making the link with QCD}
\label{sec2}
\begin{figure}
\begin{centering}
\includegraphics[scale=0.32]{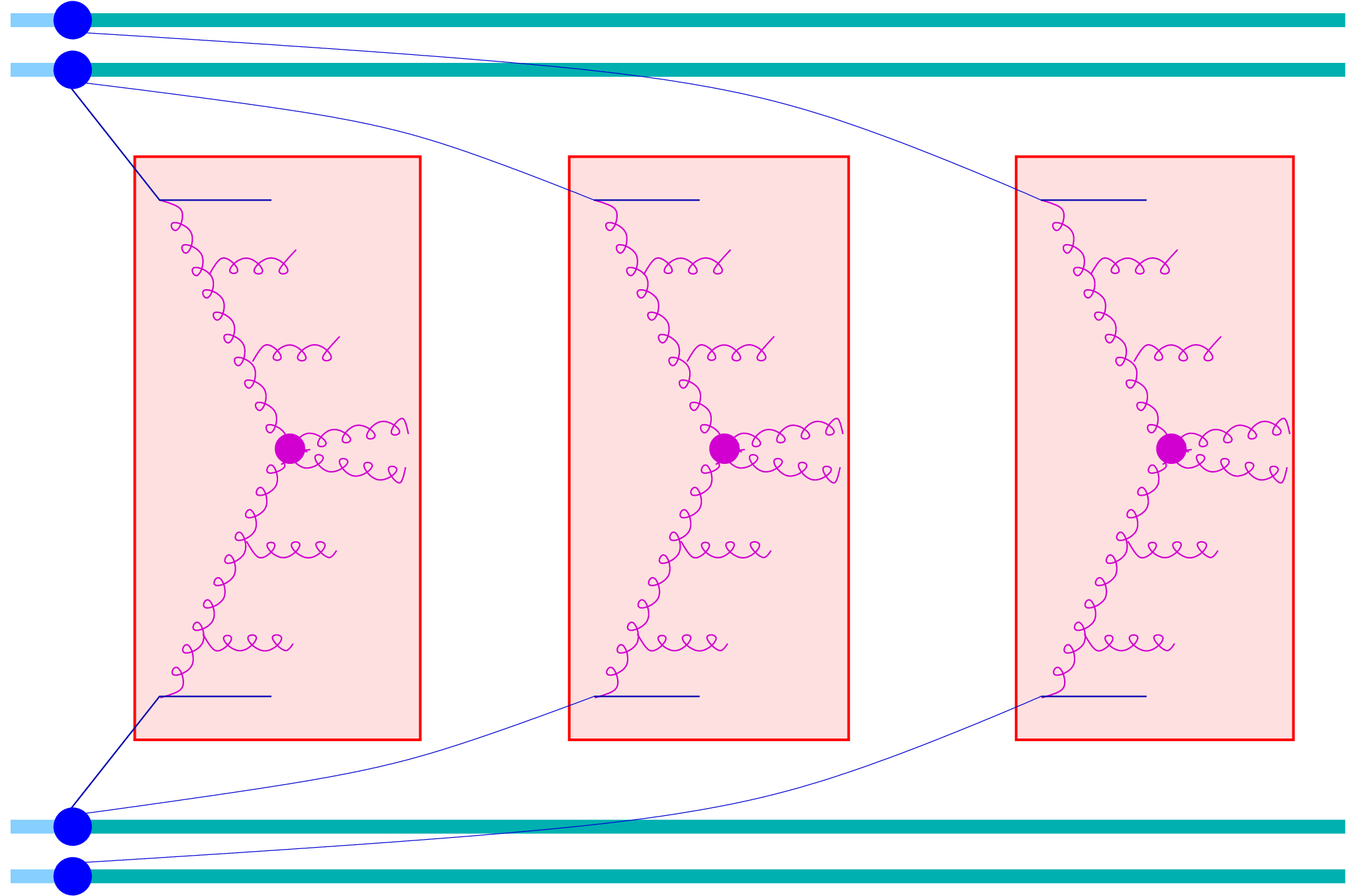}
\par\end{centering}
\caption{Using $G=G_{\mathrm{QCD}}$. \label{fig:Using-G=00003DG_QCD}}
\end{figure}

The current framework is founded on "some \emph{G}" where \emph{G} denotes a subscattering. The next step involves establishing the connection with QCD. It is assumed that 
\begin{equation}
G=G_{\mathrm{QCD}}\ ,
\end{equation}
with $G_{\mathrm{QCD}}$
denoting parton-parton scattering based on pQCD, incorporating DGLAP evolution.
(see Ref. \cite{werner:2023-epos4-heavy} and early work
(no heavy flavor) in Ref. \cite{Drescher:2000ha}). This means one
replaces the boxes of the GR approach with QCD diagrams, as sketched
in Fig. \ref{fig:Using-G=00003DG_QCD} for a collision of two nuclei
with three subscatterings. 

\begin{figure}
\begin{centering}
\includegraphics[bb=40bp 52bp 590bp 790bp,clip,scale=0.65]{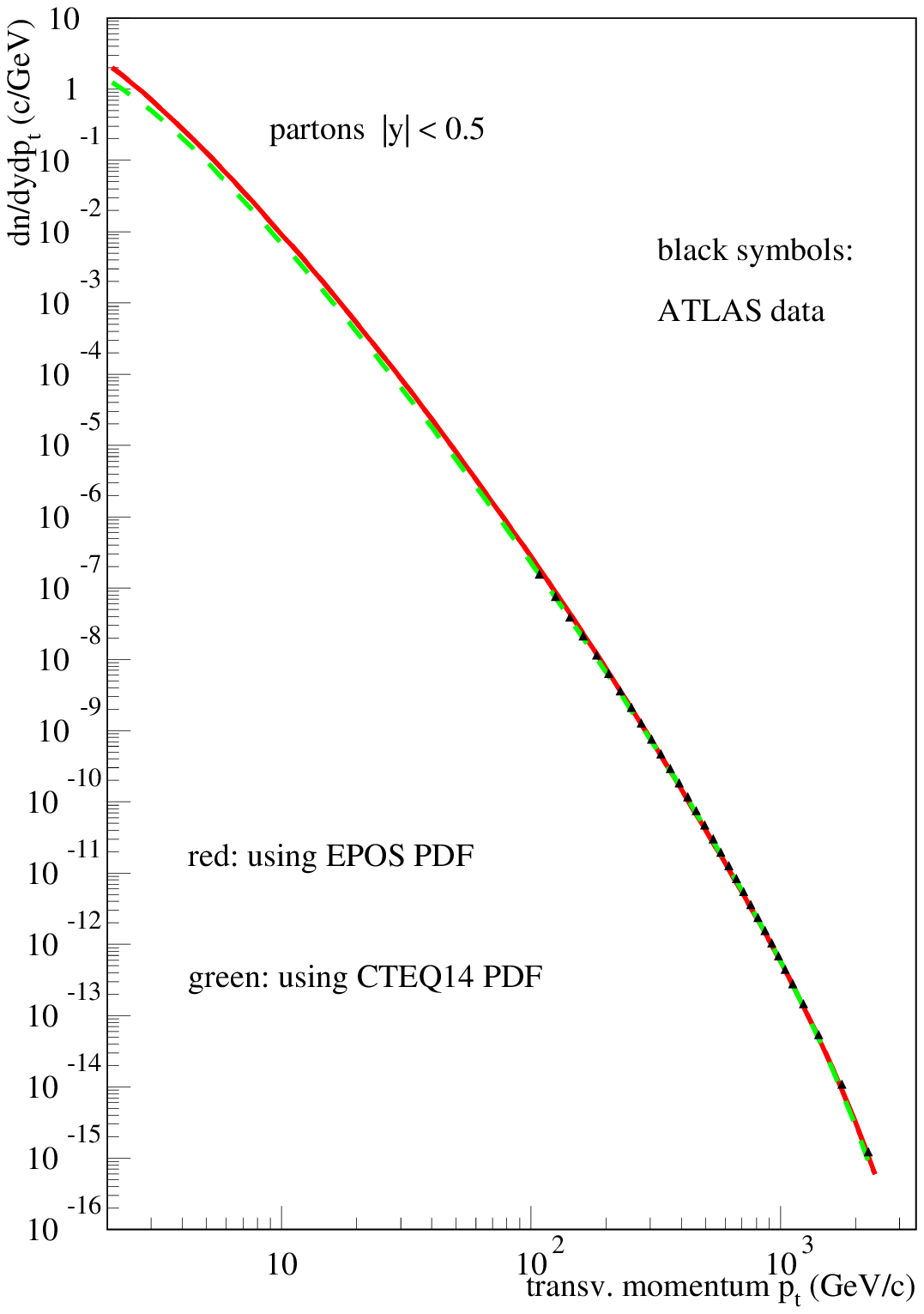}
\par\end{centering}
\caption{Jet cross section versus $p_{t}$ for pp at 13 TeV. The red line represents the EPOS result, compared to ATLAS data \cite{ATLAS:2017ble} (triangles) as well as to results derived from CTEQ PDFs
\cite{Dulat_2016-CTEQ-PDF}
(green dashed line).
The EPOS4 curve is based on EPOS4 parton distribution functions.
\label{fig:Jet-cross-section}}
\end{figure}

One calculates and tabulates "modules" (QCD evolution, Born cross sections, vertices), which enables one to assess the diagram. Various methods exist for reorganizing the modules, and one option is to establish (and tabulate) a parton distribution function (PDF), enabling the computation of the jet cross section versus $p_{t}$ for pp at 13 TeV (see Ref. \cite{werner:2023-epos4-heavy})
as shown in Fig. \ref{fig:Jet-cross-section}.
The red line represents the EPOS result, and it is compared to  ATLAS data \cite{ATLAS:2017ble} (triangles) as well as results derived from CTEQ PDFs
\cite{Dulat_2016-CTEQ-PDF}
(green dashed line).
It seems that everything is under control, but here one considered just one
single subscattering. 

In the Gribov-Regge approach, the full multiple scattering scenario
is (up to a factor $AB)$ equal to the single one for inclusive cross
sections (AGK theorem), i.e., 
\begin{equation}
\frac{d\sigma_{\mathrm{incl}}^{AB}}{dp_{t}}\ \Bigg/\ AB\times\frac{d\sigma_{\mathrm{incl}}^{\mathrm{single\,scattering}}}{dp_{t}}
\end{equation}
is unity. Unfortunately, as shown in \cite{werner:2023-epos4-smatrix},
one gets at high $p_{t}$ for this ratio 0.2 and 0.5 for minimum bias
PbPb at 5.02 TeV and $pp$ at 5.02 TeV, respectively. When trying
to understand the origin of this failure, one finds immediately (see Ref. \cite{werner:2023-epos4-smatrix}) that
it is related to energy-momentum sharing among subscatterings. Inclusive
particle spectra (like $p_{t}$ distributions) are determined by the
distribution of the LC momenta $x^{+}$ and $x^{-}$ of the subscatterings.
The squared CMS energy fraction 
\begin{equation}
x_{\mathrm{PE}}=x^{+}x^{-}\approx s\,/\,s_{\mathrm{tot}}
\end{equation}
is a crucial element, and I will explain below how the distribution of the variable $x_{\mathrm{PE}}$ is impacted by energy-momentum sharing.

\begin{figure}
\begin{centering}
\includegraphics[scale=0.70]{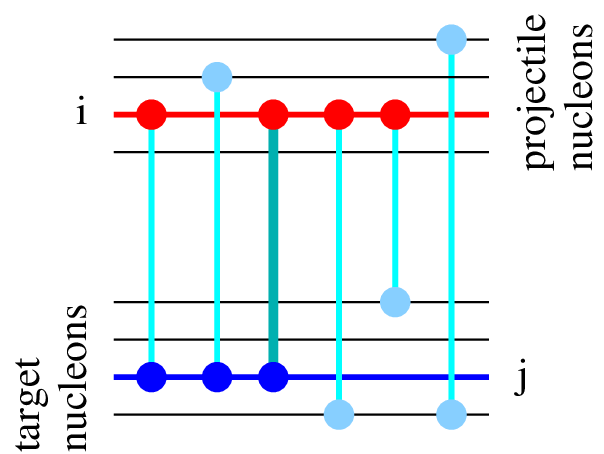}
\par\end{centering}
\caption{
For a given subscattering (Pomeron), connected to projectile nucleon \emph{i} and target nucleon \emph{j}, one defines 
the connection number $N_{\mathrm{conn}}=\frac{N_{\mathrm{P}}+N_{\mathrm{T}}}{2}$
where $N_{\mathrm{P}}$ is the number of scatterings involving nucleon
\emph{i}, and $N_{\mathrm{T}}$ the number of scatterings involving
nucleon \emph{j}.
\label{fig:The-connections-number-1}
}
\end{figure}

For a given subscattering (Pomeron), involving projectile nucleon \emph{i} and
target nucleon \emph{j,} one defines the connection number 
\begin{equation}
N_{\mathrm{conn}}=\frac{N_{\mathrm{P}}+N_{\mathrm{T}}}{2}\ ,
\end{equation}
where $N_{\mathrm{P}}$ is the number of scatterings involving nucleon
\emph{i}, and $N_{\mathrm{T}}$ the number of scatterings involving
nucleon \emph{j}, see Fig. \ref{fig:The-connections-number-1}.

\begin{figure}
\begin{centering}
\includegraphics[bb=30bp 30bp 595bp 400bp,clip,scale=0.40]{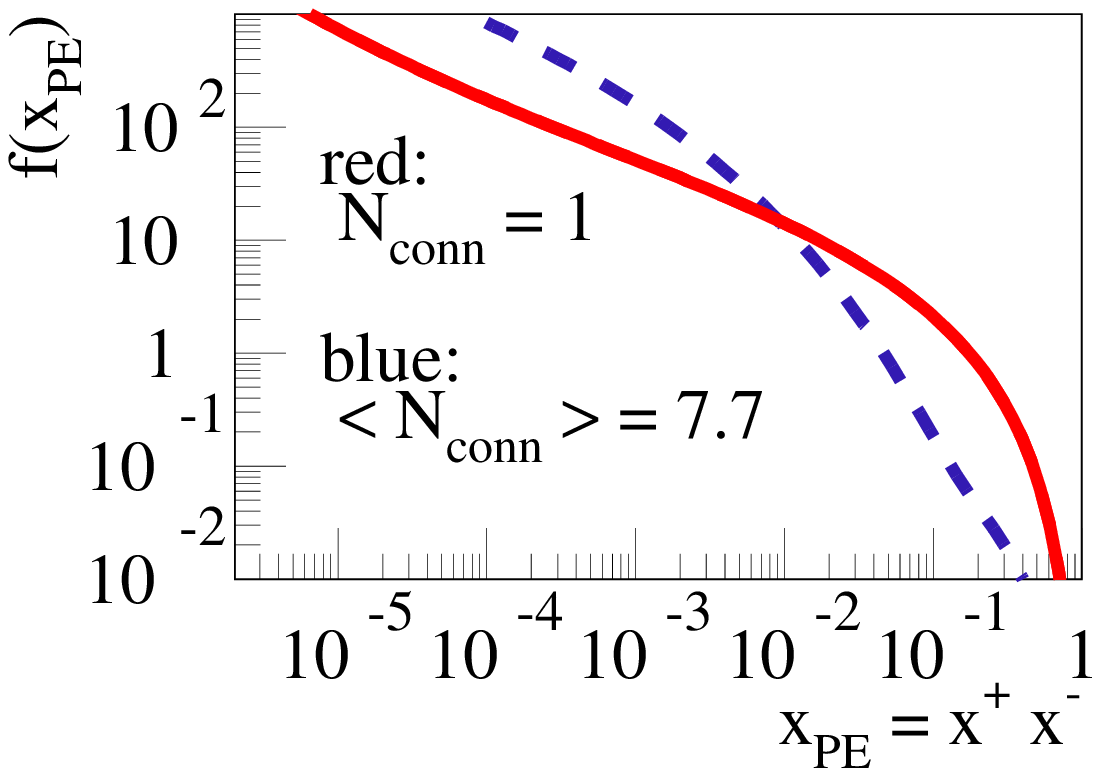}
\par\end{centering}
\caption{
The $x_{\mathrm{PE}}$ distribution $f(x_{\mathrm{PE}})$.
The red curve refers to $N_{\mathrm{conn}}=1$ (an isolated Pomeron), whereas the blue dashed one refers to central collisions with an average  $N_{\mathrm{conn}}=1$  of around 7.7.  
Large $N_{\mathrm{conn}}$ amounts unavoidably to large $x_{\mathrm{PE}}$ being suppressed, due to energy-momentum-sharing.
\label{fig:The-connections-number-2}
}
\end{figure}

The $x_{\mathrm{PE}}$ distributions $f(x_{\mathrm{PE}})$ depend
on $N_{\mathrm{conn}}$. Large $N_{\mathrm{conn}}$ amounts unavoidably
to large $x_{\mathrm{PE}}$ being suppressed, whereas small $x_{\mathrm{PE}}$
is enhanced, as shown in Fig. \ref{fig:The-connections-number-2}.
I will use the notation $f^{(N_{\mathrm{conn}})}(x_{\mathrm{PE}})$.

\begin{figure}[h]
\begin{centering}
\includegraphics[bb=30bp 20bp 576bp 432bp,clip,scale=0.45]{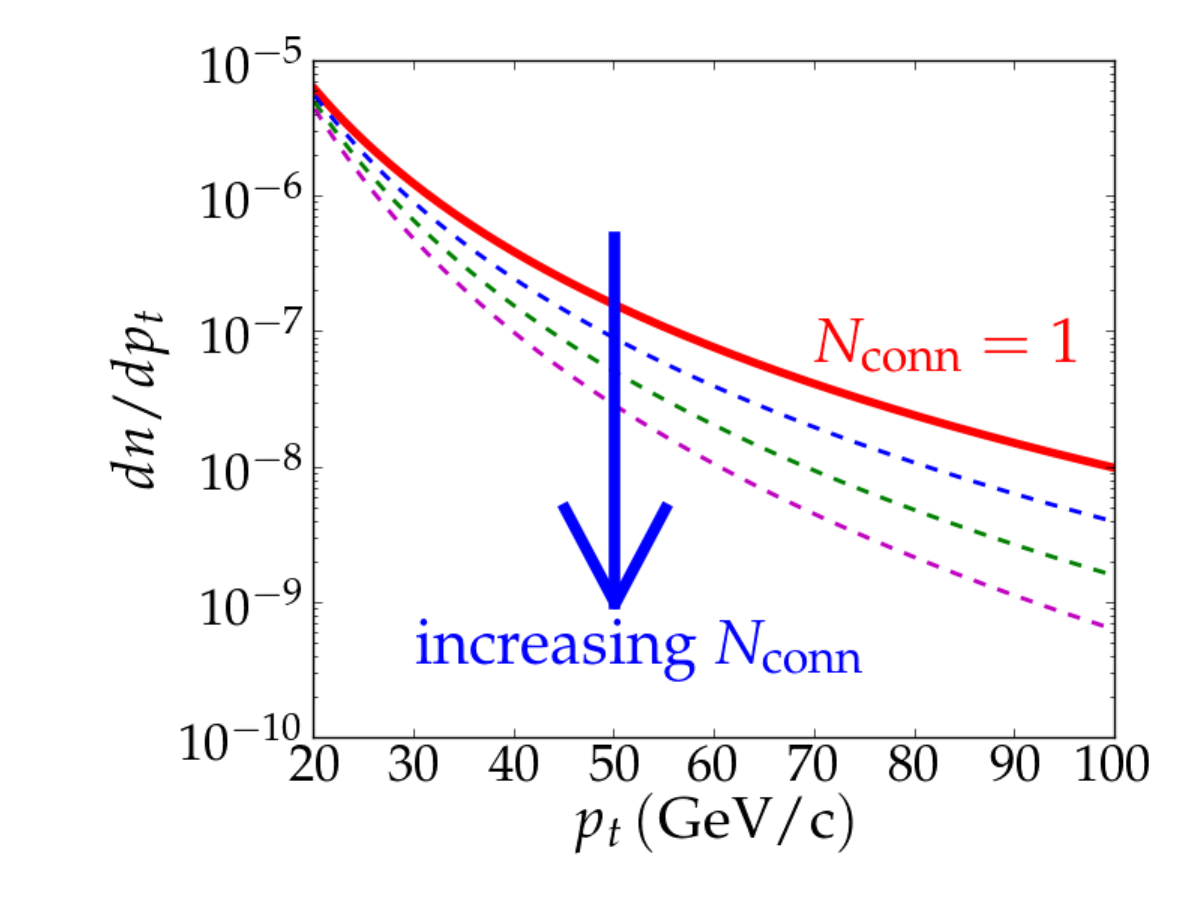}\\
\par\end{centering}
\caption{
The suppression of large values of $x_{\mathrm{PE}}$  as a consequence
of large $N_{\mathrm{conn}}$ implies a suppression of large $p_{t}$.
\label{fig:The-connections-number-3}
}
\end{figure}
\begin{figure}
\begin{centering}
\includegraphics[bb=0bp 0bp 595bp 632bp,clip,scale=0.40]{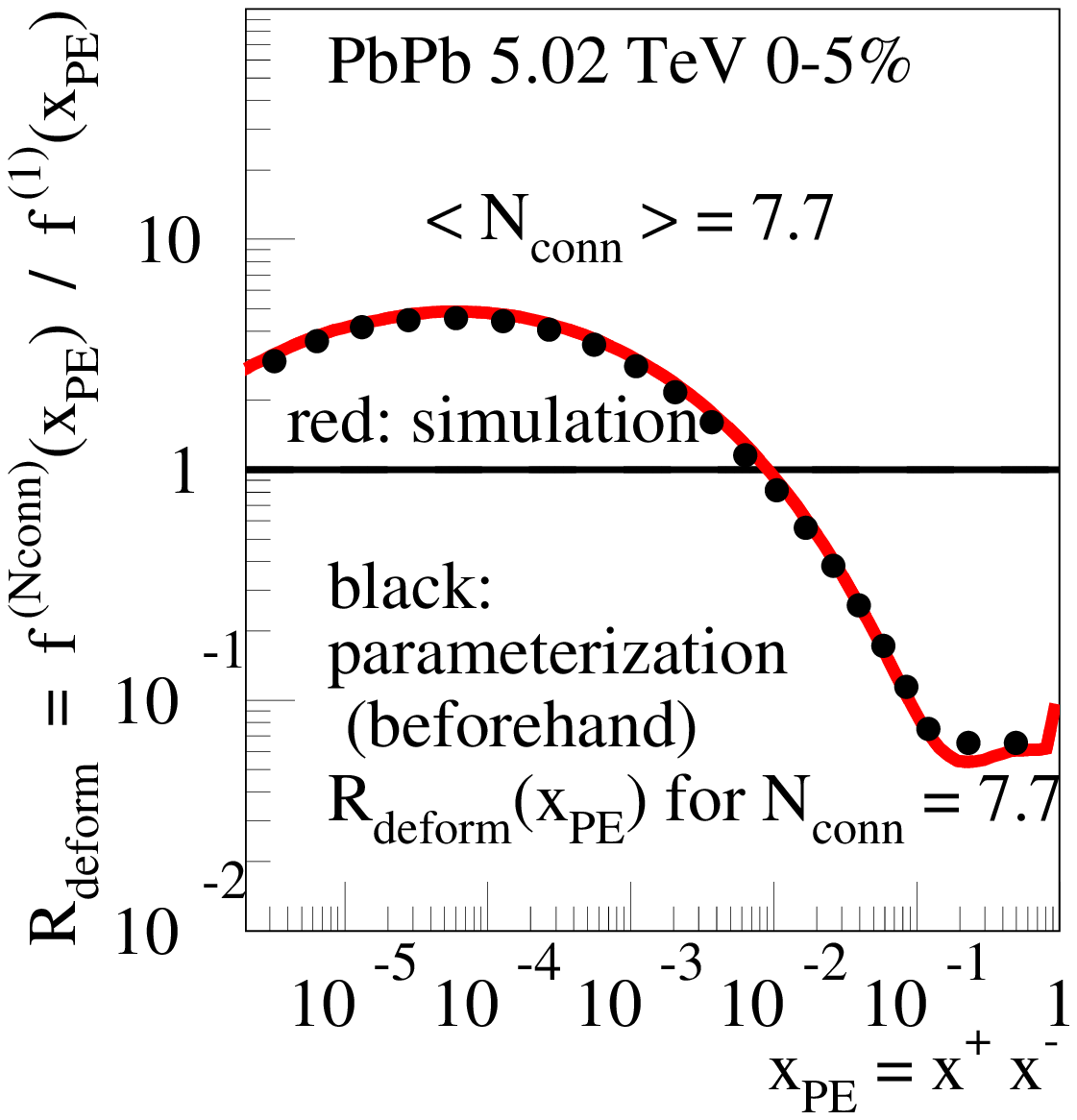}
\par\end{centering}
\caption{
The deformation function, representing the change of $f^{(N_{\mathrm{conn}})}(x_{\mathrm{PE}})$ relative to the reference
$f^{(1)}(x_{\mathrm{PE}})$. The red line corresponds to a simulation and the dotted one
to a parameterization.
\label{fig:The-connections-number-4}
}
\end{figure}

The value of $x_{\mathrm{PE}}$ is strongly correlated with the transverse momentum $p_t$ of produced particles. Pomerons with large 
 $x_{\mathrm{PE}}$ produce with a higher probability high $p_t$ particles compared to Pomerons with small $x_{\mathrm{PE}}$. For very small $x_{\mathrm{PE}}$, the hard scattering even disappears completely, and soft Pomerons take over, to produce low $p_t$ hadrons.
 
Therefore, a suppression of large values of $x_{\mathrm{PE}}$ (as a consequence
of large $N_{\mathrm{conn}}$) implies a suppression of large $p_{t}$,
as sketched in Fig. \ref{fig:The-connections-number-3}. This is
in particular true for the large $N_{\mathrm{conn}}$ contributions
in minimum bias $pp$ or $AA$ scattering. So the superposition of
the different contributions (of different values of $N_{\mathrm{conn}}$) cannot be equal to the single-scattering
case ($N_{\mathrm{conn}}=1$), one gets always a suppression at large
$p_{t}$ (and therefore a violation of AGK). \vspace{0.1cm}

As a first step towards a solution, the problem will be  "quantified".
One defines the ``deformation''
of $f^{(N_{\mathrm{conn}})}(x_{\mathrm{PE}})$ relative to the reference
$f^{(1)}(x_{\mathrm{PE}})$ as 
\begin{equation}
R_{\mathrm{deform}}=\frac{f^{(N_{\mathrm{conn}})}(x_{\mathrm{PE}})}{f^{(1)}(x_{\mathrm{PE}})}.
\end{equation}
It is $R_{\mathrm{deform}}\neq1$ which creates the problem. But one
is able to parameterize $R_{\mathrm{deform}}$ and tabulate it, for
all systems, all centrality classes (see Ref. \cite{werner:2023-epos4-smatrix}).
So 
\begin{equation}
R_{\mathrm{deform}}=R_{\mathrm{deform}}(N_{\mathrm{conn}},x_{\mathrm{PE}})
\end{equation}
can be considered to be known, it is tabulated and available via interpolation
(to be used later), see Fig. \ref{fig:The-connections-number-4},
where the red line corresponds to a simulation and the dotted one
to a parameterization.\vspace{0.1cm}

This "parameterization of the problem" will be the key element of the solution, to be sketched in the following,
and  discussed in detail in Ref. \cite{werner:2023-epos4-smatrix}.

\section{Adding saturation}

\begin{figure}
\begin{centering}
\includegraphics[scale=0.40]{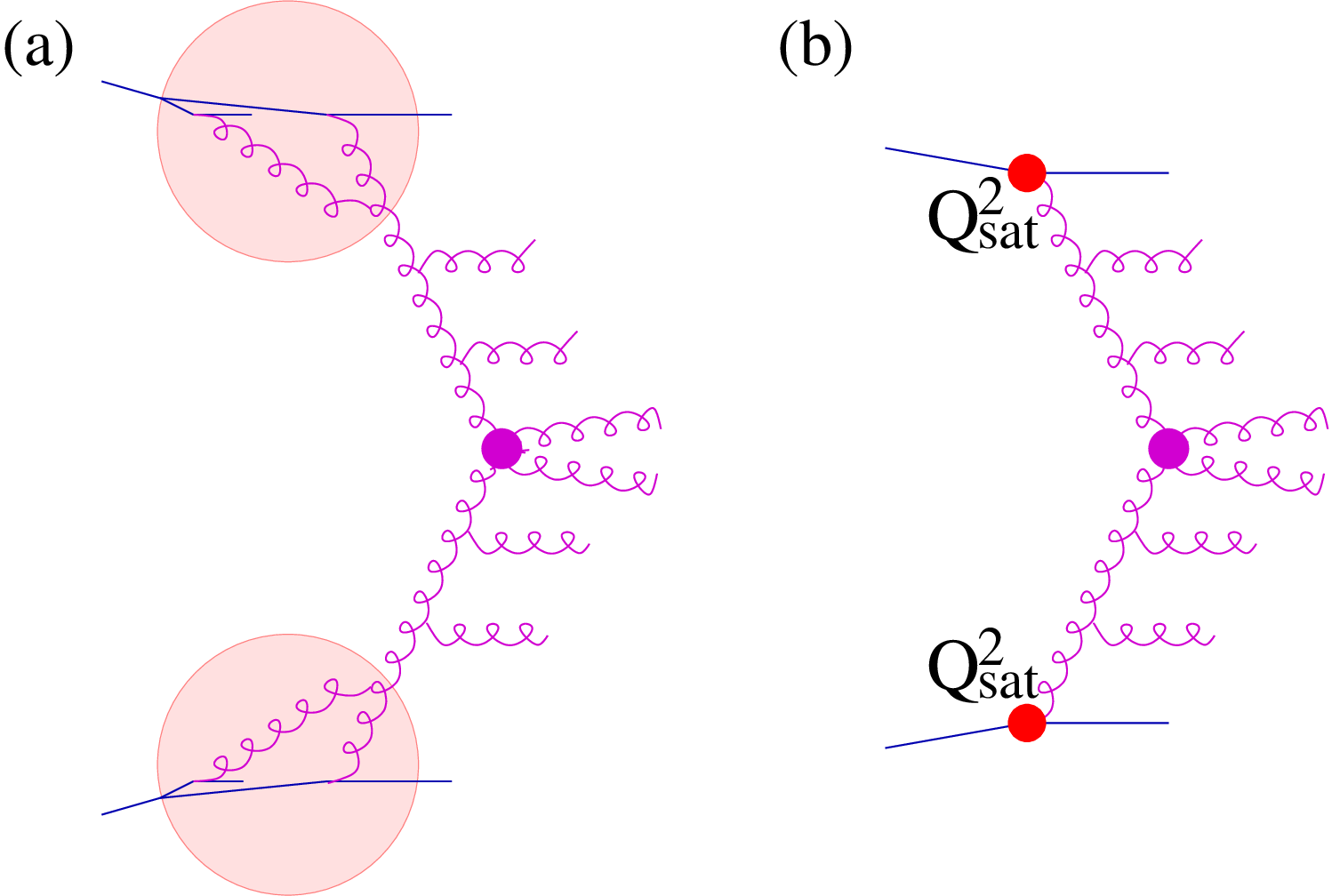}
\par\end{centering}
\caption{(a) Nonlinear effects, like gluon fusion (inside the red circles) is absent for the moment. (b) Adding nonlinear effects by introducing saturation scales $Q_{\mathrm{sat}}^{2}$ which are meant to ``summarize'' these nonlinear effects.\label{fig:Saturation-phenomena}}
\end{figure}

The single scattering expression $G$, which is the basic component of the multiple scattering formalism, actually presents two issues: (i) The assumption that $G=G_{\mathrm{QCD}}$ appears to be incorrect (AGK problem), and (ii) there is a complete absence of nonlinear effects, as shown in Fig. \ref{fig:Saturation-phenomena} (a).

The third EPOS4 improvement amounts to adding saturation, by assuming
that the nonlinear effects, inside the circles in Fig. \ref{fig:Saturation-phenomena}
(a), may be ``summarized'' by saturation scales, suggesting to
treat nonlinear effects by introducing saturation scales $Q_{\mathrm{sat}}^{2}$
as the lower limits $Q_{0}^{2}$ of the virtualities for DGLAP evolutions,
see Fig. \ref{fig:Saturation-phenomena} (b). One computes and
tabulates $G_{\mathrm{QCD}}(Q_{0}^{2},\,x^{+},x^{-},s,b)$ for a large
range of $Q_{0}^{2}$ values, see Ref. \cite{werner:2023-epos4-heavy}. 

\begin{figure}
\begin{centering}
\includegraphics[scale=0.40]{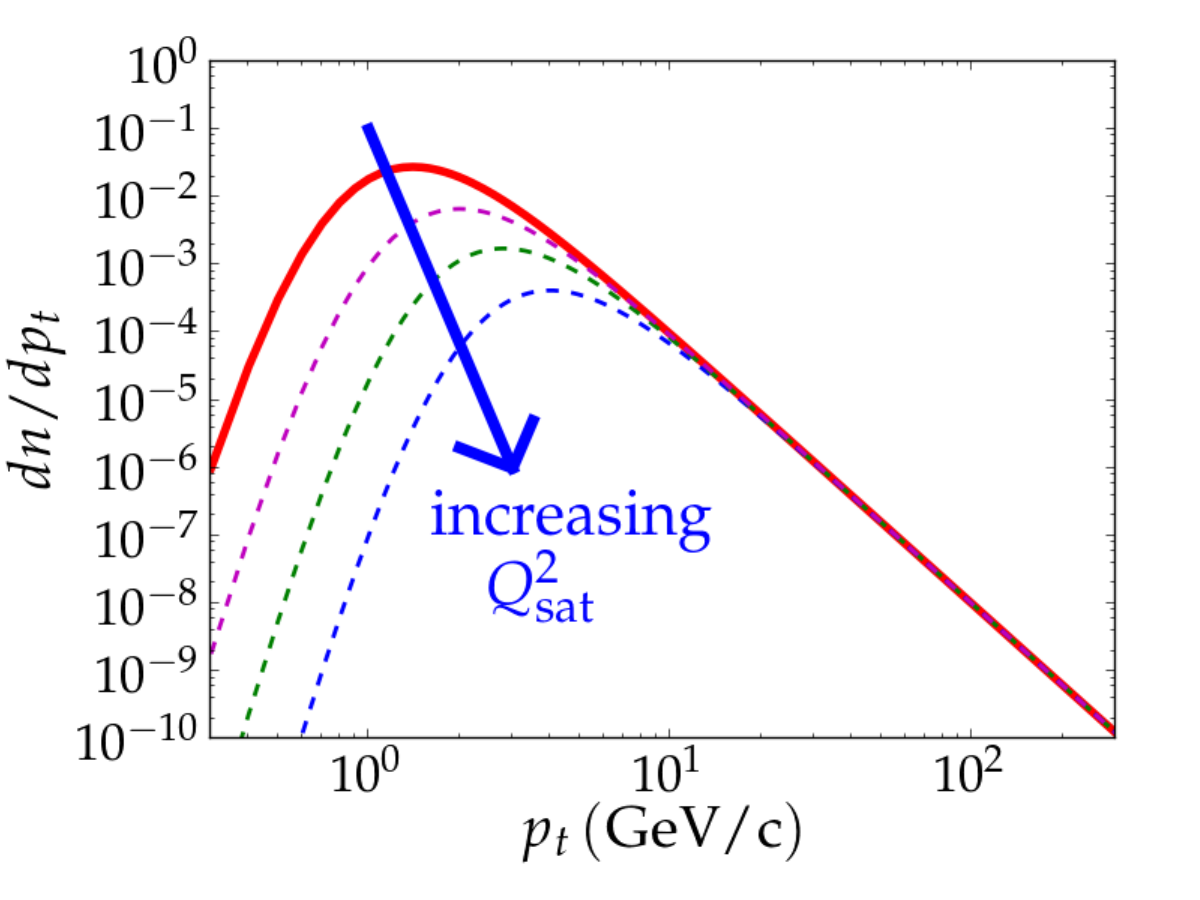}
\par\end{centering}
\caption{Contributions with increasing $N_{\mathrm{conn}}$. \label{fig:Contributions-with-increasing} }
\end{figure}

Concerning the connection between the basic multiple scattering building
block $G$ and the QCD expression $G_{\mathrm{QCD}}$, one postulates
that for each subscattering, for given $x^{\pm}$, $s$, $b$, and
$N_{\mathrm{conn}}$, one has 
\begin{equation}
G(\,x^{+},x^{-},s,b)=n\frac{G_{\mathrm{QCD}}(Q_{\mathrm{sat}}^{2}\,,\,x^{+},x^{-},s,b)}{R_{\mathrm{deform}}(N_{\mathrm{conn}},x_{\mathrm{PE}})},\label{basic-epos4-equation}
\end{equation}
\textbf{such that $\boldsymbol{G}$ does not depend on $\boldsymbol{N_{\mathrm{conn}}}$},
whereas $Q_{\mathrm{sat}}^{2}$ does so. Here, $n$ is a normalization
constant. Using  Eq. (\ref{basic-epos4-equation}), one can show \cite{werner:2023-epos4-smatrix}:
\begin{equation}
\frac{d^{2}\sigma_{\mathrm{incl}}^{AB\,(N_{\mathrm{conn}})}}{dx^{+}dx^{-}}\propto\frac{d\sigma_{\mathrm{incl}}^{\mathrm{single\,scattering}}}{dx^{+}dx^{-}}\left[Q_{\mathrm{sat}}^{2}(N_{\mathrm{conn}},x^{+},x^{-})\right],
\end{equation}
i.e., the A+B cross section (for given given $N_{\mathrm{conn}}$)
is equal to the single scattering case, but with $Q_{\mathrm{sat}}^{2}$
corresponding to $N_{\mathrm{conn}}$. The same relation holds for
p\textsubscript{\emph{t}} distributions (deduced from x\textsuperscript{+}x\textsuperscript{-}).
One expects, as sketched in Fig. \ref{fig:Contributions-with-increasing},
with increasing $N_{\mathrm{conn}}$ an increasing $Q_{\mathrm{sat}}^{2}$,
and a reduction at $p_{t}^{2}<Q_{\mathrm{sat}}^{2}$ compared to $N_{\mathrm{conn}}=1$
(red curve). But no change for large p\textsubscript{\emph{t}}.

\begin{figure}
\begin{centering}
\includegraphics[bb=20bp 50bp 590bp 790bp,clip,scale=0.65]{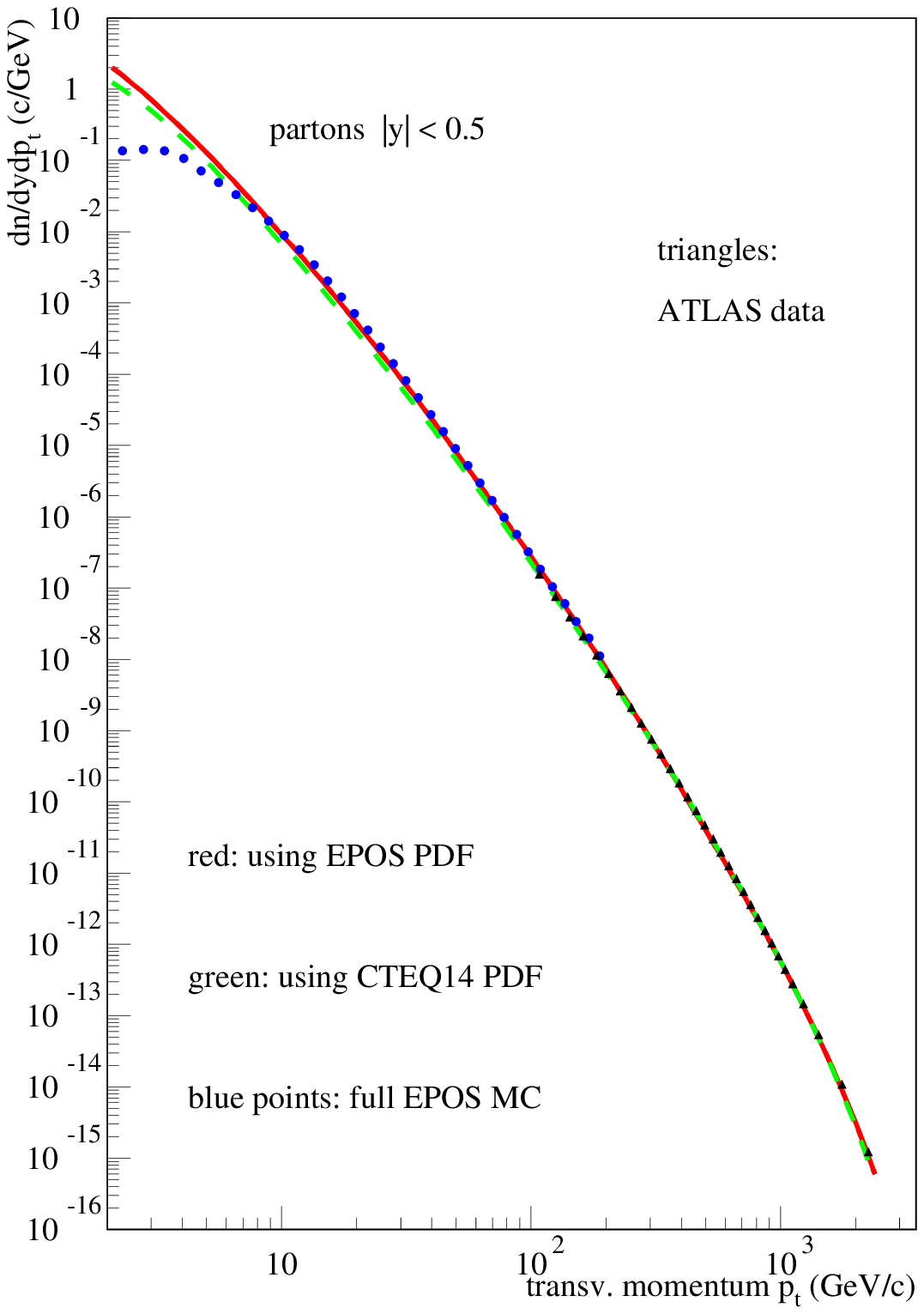}
\par\end{centering}
\caption{Jet cross section versus $p_{t}$ for pp at 13 TeV. The full (multiple scattering) Monte Carlo
result (blue points) is compared to the EPOS result for one single Pomeron (red line) and to  ATLAS data \cite{ATLAS:2017ble} (triangles). I show as well a result based on factorization, derived from CTEQ PDFs \cite{Dulat_2016-CTEQ-PDF}
(green dashed line).
\label{fig:Jet-cross-section-1} }
\end{figure}

If one is interested in large p\textsubscript{\emph{t}}, one replaces
$Q_{\mathrm{sat}}^{2}$ by some constant $Q_{0}^{2}=\max\{Q_{\mathrm{sat}}^{2}\}$,
and one gets finally
\begin{equation}
\frac{d\sigma_{\mathrm{incl}}^{AB\,(mb)}}{dp_{t}}=AB\frac{d\sigma_{\mathrm{incl}}^{\mathrm{single\,scattering}}}{dp_{t}}\left[Q_{0}^{2}\right],
\end{equation}
but only for $p_{t}^{2}$ bigger than the relevant $Q_{\mathrm{sat}}^{2}$
values (a kind of generalized AGK theorem). This is extremely important:
one gets (for the first time) factorization (in \emph{pp} and \emph{A}+\emph{B})
for inclusive cross sections at high $p_{t}$ in a fully selfconsistent\footnote{\noindent Mandatory: (A) energy-momentum conservation, (B) parallel
scattering, (C) MC = theory, (D) factorization for high $p_{t}$ } multiple (parallel) scattering scheme. What this means, is shown
in Fig. \ref{fig:Jet-cross-section-1}, where the jet cross section
for $pp$ at 13 TeV is plotted. This is the same plot
as shown earlier, but here I add in addition the full Monte Carlo
result (blue points). Important: The Monte Carlo
curve agrees at large $p_t$ with the red curve, representing the single
Pomeron result based on PDFs. \textbf{Without the requirement formulated
in Eq. (\ref{basic-epos4-equation}), the Monte Carlo result (blue)
would be a factor 5 below the single Pomeron case (red). }

\begin{figure}
\begin{centering}
\includegraphics[scale=0.32]{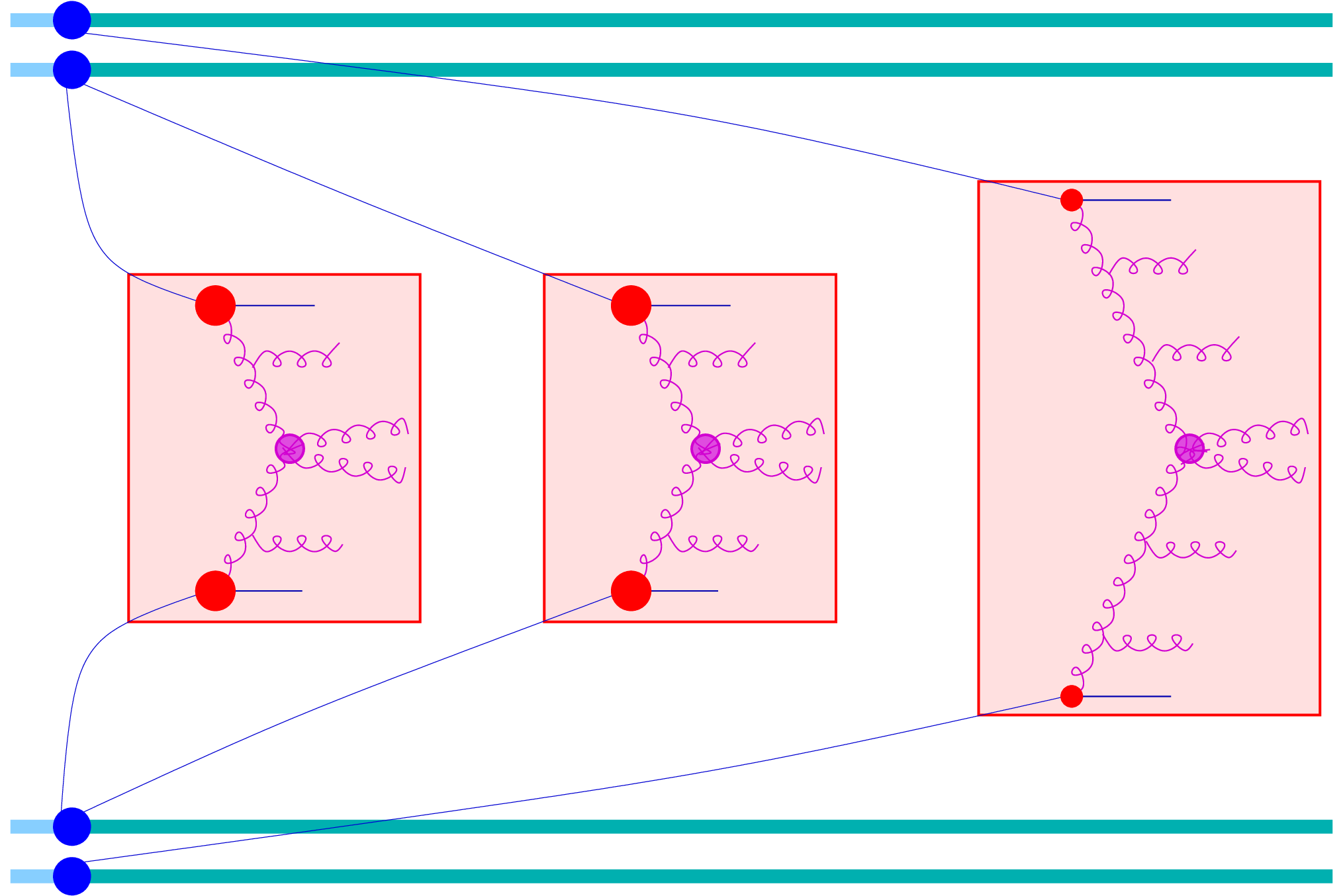}
\par\end{centering}
\caption{Sketch of the ``compensation'' of smaller energies (red box sizes)
by larger saturation scale values (red dots).\label{fig:Compensation-of-the}}
\end{figure}

How to understand why these $N_{\mathrm{conn}}$-dependent saturation
scales ``work''? Let me qualitatively explain this by
considering an $A+B$ scattering ($A=B=2$) with 3 subscatterings,
as shown in Fig. \ref{fig:Compensation-of-the} (more quantitative
discussions can be found in Refs. \cite{werner:2023-epos4-heavy,werner:2023-epos4-smatrix}).
Let me consider any of the two left scatterings, compared to the right
one: $N_{\mathrm{conn}}$ is bigger (2 compared to 1); the energy
($\sqrt{s}$) smaller due to energy sharing; $Q_{\mathrm{sat}}^{2}$
is bigger because of the larger $N_{\mathrm{conn}}$~(bigger~dots);
the parton evolution shorter due to the bigger $Q_{\mathrm{sat}}^{2}$;
\textbf{the central part responsible for the hard scattering is identical.}
This last point is the crucial element, which assures that at the
end the hard particle production is identical independent of $N_{\mathrm{conn}}$
, and therefore the sum of all $N_{\mathrm{conn}}$ contributions
is (up to a factor) identical to the single Pomeron case. And this
is what is needed to get factorization in such a multiple scattering
formalism.

\section{Conclusion}

I explained the new concepts underpinning EPOS4. Starting from the GR approach (which ensures already parallel scatterings), I explain the improvements in three steps: (a) adding energy-momentum conservation; (b) making the link with QCD; (c) adding saturation. The latter is done such that the (unavoidable) deformation of the Pomerons energy distribution in the case of many parallel scatterings, is completely "absorbed" into the saturation scale, which ensures that at the end the high $p_t$ values are not affected. This is nothing less than a reconciliation of multiple scattering, featuring parallel scatterings, and factorization.
\bibliographystyle{SciPost_bibstyle}

\begin{thebibliography}{10}
\providecommand{\url}[1]{\texttt{#1}}
\providecommand{\urlprefix}{URL }
\expandafter\ifx\csname urlstyle\endcsname\relax
  \providecommand{\doi}[1]{doi:\discretionary{}{}{}#1}\else
  \providecommand{\doi}{doi:\discretionary{}{}{}\begingroup
  \urlstyle{rm}\Url}\fi
\providecommand{\eprint}[2][]{\url{#2}}

\bibitem{werner:2023-epos4-overview}
K.~Werner,
\newblock \emph{Revealing a deep connection between factorization and
  saturation: New insight into modeling high-energy proton-proton and
  nucleus-nucleus scattering in the epos4 framework},
\newblock Phys. Rev. C \textbf{108}, 064903 (2023),
\newblock \doi{10.1103/PhysRevC.108.064903}.

\bibitem{werner:2023-epos4-heavy}
K.~Werner and B.~Guiot,
\newblock \emph{{Perturbative QCD concerning light and heavy flavor in the
  EPOS4 framework}},
\newblock Phys. Rev. C \textbf{108}, 034904 (2023),
\newblock \doi{10.1103/PhysRevC.108.034904}.

\bibitem{werner:2023-epos4-smatrix}
K.~Werner,
\newblock \emph{{Parallel scattering, saturation, and generalized
  Abramovskii-Gribov-Kancheli (AGK) theorem in the EPOS4 framework, with
  applications for heavy-ion collisions at sNN of 5.02 TeV and 200 GeV}},
\newblock Phys. Rev. C \textbf{109}(3), 034918 (2024),
\newblock \doi{10.1103/PhysRevC.109.034918}.

\bibitem{werner:2023-epos4-micro}
K.~Werner,
\newblock \emph{Core-corona procedure and microcanonical hadronization to
  understand strangeness enhancement in proton-proton and heavy ion collisions
  in the epos4 framework},
\newblock Phys. Rev. C \textbf{109}, 014910 (2024),
\newblock \doi{10.1103/PhysRevC.109.014910}.

\bibitem{Gribov:1967vfb}
V.~N. Gribov,
\newblock \emph{{A REGGEON DIAGRAM TECHNIQUE}},
\newblock Zh. Eksp. Teor. Fiz. \textbf{53}, 654 (1967).

\bibitem{Gribov:1968jf}
V.~N. Gribov,
\newblock \emph{{Glauber corrections and the interaction between high-energy
  hadrons and nuclei}},
\newblock Sov. Phys. JETP \textbf{29}, 483 (1969).

\bibitem{GribovLipatov:1972}
V.~N. Gribov and L.~N. Lipatov,
\newblock \emph{{}},
\newblock Sov. J. Nucl. Phys. \textbf{15}, 438 (1972).

\bibitem{Abramovskii:1973fm}
V.~A. Abramovskii, V.~N. Gribov and O.~V. Kancheli,
\newblock \emph{{Character of Inclusive Spectra and Fluctuations Produced in
  Inelastic Processes by Multi - Pomeron Exchange}},
\newblock Yad. Fiz. \textbf{18}, 595 (1973).

\bibitem{Collins:1989}
J.~Collins, D.~Soper and G.~Sterman,
\newblock \emph{{}},
\newblock in Perturbative Quantum Chromodynamics, edited by A.H. Mueller, World
  Scientific, Singapore  (1989).

\bibitem{Ellis:1996}
R.~Ellis, W.~Stirling and B.~Webber,
\newblock \emph{{}},
\newblock in QCD and Collider Physics, Cambridge Monographs on Particle
  Physics, Nuclear Physics and Cosmology  (1996).

\bibitem{Drescher:2000ha}
H.~J. Drescher, M.~Hladik, S.~Ostapchenko, T.~Pierog and K.~Werner,
\newblock \emph{{Parton based Gribov-Regge theory}},
\newblock Phys. Rep. \textbf{350}, 93 (2001),
\newblock \doi{10.1016/S0370-1573(00)00122-8}.

\bibitem{ATLAS:2017ble}
M.~Aaboud \emph{et~al.},
\newblock \emph{{Measurement of inclusive jet and dijet cross-sections in
  proton-proton collisions at $\sqrt{s}=13$ TeV with the ATLAS detector}},
\newblock JHEP \textbf{05}, 195 (2018),
\newblock \doi{10.1007/JHEP05(2018)195}.

\bibitem{Dulat_2016-CTEQ-PDF}
S.~Dulat, T.-J. Hou, J.~Gao, M.~Guzzi, J.~Huston, P.~Nadolsky, J.~Pumplin,
  C.~Schmidt, D.~Stump and C.-P. Yuan,
\newblock \emph{New parton distribution functions from a global analysis of
  quantum chromodynamics},
\newblock Physical Review D \textbf{93}(3) (2016),
\newblock \doi{10.1103/physrevd.93.033006}.

\end{thebibliography}


\end{document}